\documentstyle[cite,t1enc,graphics,prettyref,eqsecnum,preprint,tighten,aps,amstex]{revtex}

\makeatletter

\providecommand{\LyX}{L\kern-.1667em\lower.25em\hbox{Y}\kern-.125emX\@}

\makeatother

\begin{document}

\preprint{\begin{tabular}{r} FTUV--00--0815\\ IFIC/00--48 \end{tabular}}

\draft

\title{Extra dimensions at the one loop level:\\
\( Z\rightarrow b\bar{b} \) and \( B \)-\( \bar{B} \) mixing}

\author{Joannis Papavassiliou and Arcadi Santamaria}

\address{\hfill{}\\
 Departament de Física Teòrica and IFIC, Universitat de València -- CSIC\\
 Dr. Moliner 50, E-46100 Burjassot (València), Spain}

\maketitle
\begin{abstract}
We study, at the one loop level, the dominant new physics contributions from
extra dimensions to \( Z\rightarrow b\bar{b} \), as well as \( B \)-\( \bar{B} \)
and \( K \)-\( \bar{K} \) mixing. We use a model with one extra dimension
containing fermions which live in four dimensions, and gauge bosons and one
scalar doublet propagating in five dimensions. We find that the effect of the
infinite tower of Kaluza-Klein modes in \( Z\rightarrow b\bar{b} \) is finite
and gives a negative correction to \( R_{b}=\Gamma _{b}/\Gamma _{h} \), which
is used to set a lower bound of 1 TeV on the compactification scale \( M_{c} \).
On the other hand, we show that the box diagrams contributing to \( B \)-\( \bar{B} \)
and \( K \)-\( \bar{K} \) mixing are divergent and, after proper regularization,
we find that they increase the value of the function \( S(x_{t}) \) which governs
this mixing. The obtained value is perfectly compatible with available data. 
\end{abstract}

\section{Introduction }

\label{sec:intro}In the last years there has been a revival of interest in
new physics scenarios in which the ordinary four dimensional standard model
(SM) arises as a low energy effective theory of models defined in five or more
dimensions. Apart from the fact that this type of models arise naturally in
string scenarios, there are various reasons for this renewed interest. Probably
the most exciting one is the realization that the size of the extra dimensions
can be amazingly large without contradicting present experimental data \cite{arkani-hamed:1998nn,arkani-hamed:1998rs,antoniadis:1990ew,antoniadis:1994jp}.
This opens the door to the possibility of testing these models in the near future.
In fact, a general feature of models with large extra dimensions is the presence
of a tower of Kaluza-Klein (KK) states which, if light enough, could be produced
in the next generation of accelerators (see for instance \cite{antoniadis:1994yi,antoniadis:1999bq,nath:1999mw}).
In addition, models based on large extra dimensions can be used to shed light
on a variety of problems. First of all, by introducing a new scale close to
the electroweak scale the hierarchy problem is pushed by a few orders of magnitude
\cite{arkani-hamed:1998rs,arkani-hamed:1998kx}. Furthermore, by resorting to
extra dimensions one might gain new insights on the size of the cosmological
constant \cite{arkani-hamed:2000eg,sundrum:1997js}. In addition, supersymmetry
breaking could be explained in the context of such theories \cite{antoniadis:1993fh}.
Moreover, the linear running of gauge couplings obtained in models with extra
dimensions can be used to lower the scale of gauge coupling unification (see
for instance \cite{dienes:1998vg,kakushadze:1998vr}). Finally, by assigning
fermions to different configurations of the extra dimensions one hopes to reproduce
the hierarchical pattern of fermion masses (see for instance \cite{bando:2000it,arkani-hamed:1998vp,dienes:1998sb,dvali:1999cn,ioannisian:1999cw}). 

Models with compact extra dimensions are in general not renormalizable, and
one should regard them as low energy manifestations of some more fundamental
theory, perhaps string theory. The effects of the extra dimensions are communicated
to the four dimensional world through the presence of infinite towers of KK
modes, which modify qualitatively the behavior of the low energy theory. In
particular, the non-renormalizability of the theory is found when summing the
infinite tower of KK states. Indeed, already when computing tree level processes,
one encounters sums of the type

\begin{equation}
\sum ^{+\infty }_{n_{1},n_{2},\cdots =-\infty }\frac{1}{n^{2}_{1}+n^{2}_{2}+\cdots +n_{\delta }^{2}}\, ,
\end{equation}
where \( \delta  \) is the number of extra dimensions. The above sum is divergent
if \( \delta >1 \). Notice that this type of behavior is different from conventional
non-renormalizable theories where, at least at tree level, all processes are
finite. Then, if \( \delta >1 \) one readily assumes that the theory should
be cut off at some scale above the compactification scale. In practice this
is implemented by truncating the tower of KK modes at \( n_{i}\sim 100 \).
Such a truncation mechanism is dynamically realized in the context of some string
theories, where an exponential dumping factor suppresses the couplings of the
KK modes to ordinary matter \cite{antoniadis:1994jp}. Models with only one
extra dimension (\( \delta =1 \)) are especially interesting because the above
sum is convergent. Therefore, the tree level predictions of five dimensional
models are particularly stable with respect to the scale of any new physics
beyond the compactification scale. However, as commented before, even such models
are not renormalizable, and one expects that their bad high energy behavior
will eventually manifest itself also at the level of the four dimensional theory
with an infinity of KK modes. Thus, it is interesting to study the behavior
of this type of models at the one loop level and investigate to what extent
their good tree level behavior is maintained. We will show in section~\ref{sec:zbb}
that the effect of summing the infinite tower of KK modes amounts to changing
the propagator of the particle having KK modes by a propagator which behaves
like \( 1/k \) for large \( k \), instead of the canonical \( 1/k^{2} \)
behavior. This ultimately will trigger the non-renormalizability even of models
with only one extra dimension. In spite of that the integrals involving only
one summation over KK modes are as well behaved as their counterparts in the
original (zero-mode) renormalizable four dimensional theory; they too will therefore
give rise to finite results.

Models with extra dimensions are also interesting from the phenomenological
point of view because they are very predictive once the spectrum and the symmetries
have been specified (e.g. which fields live in four dimensions and which fields
live in the extra dimensions). For instance, five dimensional extensions of
the SM or the minimal supersymmetric standard model (MSSM) contain only one
additional parameter, the compactification radius, \( R \), or its inverse,
the compactification scale \( M_{c}=1/R \). In principle the theory also depends
on the cutoff scale of the theory \( M_{s}<100M_{c} \); however, for models
with a single extra dimension this scale does not appear at tree level and,
as we will see, many one-loop results are also rather insensitive to it. On
the other hand, models with more than one extra dimension can depend heavily
on this additional parameter. 

In this paper we study a model with only one extra dimension at the one loop
level following the bottom-up approach. Specifically, we will build a four-dimensional
quantum field theory (QFT) containing an infinite tower of KK modes, derived
from a five dimensional model. In this framework we will study some of the theoretical
issues that arise when keeping the infinite tower of KK modes, as well as some
of their phenomenological consequences. 

There are many different types of models with large extra dimensions depending
on the fields they contain and the exact location of these fields~\cite{arkani-hamed:1998rs}.
For our purposes we will adopt the simplest generalization of the SM, namely
the so-called 5DSM with fermions living in four dimensions and gauge bosons
and a single scalar doublet propagating in five dimensions\cite{pomarol:1998sd}.
This simple model will allow us to explore the behavior of the theory at the
one loop level and, at the same time, to extract some phenomenological constraints
derived from one loop processes which are enhanced due to their strong dependence
on the top-quark mass, \( m_{t} \). Thus, in section~\ref{sec:lagrangian}
we derive the relevant four dimensional Lagrangian containing the tower of KK
modes from the five dimensional one. At energies much smaller than the compactification
scale the tower of KK modes can be integrated out. This gives rise to a four-fermion
interaction, which is also derived in section~\ref{sec:lagrangian}. In section~\ref{sec:zbb}
we use the process \( Z\rightarrow b\bar{b} \) as a laboratory to study the
effect of the KK tower of charged scalar fields at the one-loop level. This
process is also phenomenologically interesting because it is very well measured
and because it is sensitive to the presence of additional scalar fields with
couplings proportional to \( m_{t} \). We find that the scalar KK modes give
rise to a finite contribution, and discuss the reason for that. The theoretical
prediction thus obtained, combined with the existing precise experimental value
of \( R_{b} \), is used to set stringent bounds on the compactification scale.
Section~\ref{sec:box} is devoted to the study of two related processes, namely
\( K \)-\( \bar{K} \) and \( B \)-\( \bar{B} \) mixing, induced by box diagrams
involving the exchange of two scalar towers of KK modes. These diagrams are
also enhanced by the top quark mass and are interesting from the phenomenological
point of view. Contrary to the case of \( Z\rightarrow b\bar{b} \), the presence
of two towers of KK modes renders these diagrams divergent. Introducing the
cutoff of the theory, \( M_{s}, \) we estimate their contribution and compare
it with the available experimental data. Finally in section~\ref{sec:conclusions}
we collect and discuss the results.

\section{The Lagrangian}

\label{sec:lagrangian}When studying the dominant radiative corrections induced
by the exchange of KK modes, it is natural to focus on processes which are known
to be sensitive to radiative corrections even in the absence of KK modes. In
the SM the most important loop effects are those enhanced due to the dependence
on the heavy top-quark mass: \( Z\rightarrow b\bar{b} \) \cite{akhundov:1986fc,bernabeu:1988me,bernabeu:1991ws,beenakker:1988pv},
\( B \)-\( \bar{B} \)-mixing \cite{buchalla:1996vs}, and the \( \rho  \)
parameter. 

If fermions live in four dimensions, as is the case in the model we consider,
there are no KK modes associated with the top quark; therefore, there are no
additional one-loop corrections to the \( \rho  \) parameter enhanced by the
top-quark mass. On the other hand, in models with gauge bosons living in the
extra dimensions the \( \rho  \) parameter is already modified at tree level,
because the KK modes of gauge bosons mix with the standard zero-mode gauge bosons,
a fact which provides interesting constraints on the compactification scale~\cite{masip:1999mk,delgado:1999sv,rizzo:1999br,carone:1999nz,nath:1999fs,nath:1999mw}.
We will therefore focus on the remaining two processes mentioned above. 

In the SM the dominant contributions to \( Z\rightarrow b\bar{b} \) and \( B \)-\( \bar{B} \)
-mixing come from diagrams with the charged scalars (the would-be Goldstone
bosons) running in the loop, because their couplings are proportional to the
top-quark mass. One can easily establish this in the Feynman or in the Landau
gauges. The contributions from the exchange of gauge bosons are suppressed by
powers of \( (m_{W}/m_{t})^{2} \) and vanish in the gauge-less limit (\( g\rightarrow 0) \)
or in the large top-quark mass limit. However, because the top quark mass is
not so heavy, the convergence of the expansion is rather slow \cite{bernabeu:1988me}
and the complete calculation is needed in order to match the experimental accuracy
against the SM prediction. In spite of that, the dominant large top-quark mass
approximation is good enough for many purposes, in particular when estimating
the size of contributions stemming from new physics. 

If the scalar doublet lives in five dimensions it will give rise to a tower
of KK modes with Yukawa couplings proportional to the top-quark mass. Therefore,
we expect the contributions from diagrams containing these couplings to be numerically
dominant. If the scalar doublet lives in four dimensions there could still be
important contributions coming from the exchange of the KK modes of the gauge
bosons, but they are not enhanced by the top-quark mass. Therefore, we will
only consider the coupling of a scalar doublet living in five dimensions to
fermions living in four dimensions.

The relevant pieces of the five dimensional Lagrangian are (\( \mu =0,1,2,3 \)
are four dimensional indices and \( M=0,1,2,3,5 \)\footnote{%
Following the standard notation we label the fifth component as \( 5 \), even
though we started at \( 0 \).
} are five dimensional ones) \begin{equation}
\label{eq:5d-lagrangian}
L=\int d^{5}x\left( \partial _{M}\varphi ^{\dagger }\partial ^{M}\varphi -\left( \bar{Q}_{L}Y_{u}u_{R}\, \varphi \, \delta (x^{5})+\mathrm{h}.\mathrm{c}.\right) +\cdots \right) \, ,
\end{equation}
 where \( \varphi (x^{M}) \) is the \( SU(2) \) scalar doublet which lives
in five dimensions. \( Q_{L}(x^{\mu }) \) and \( u_{R}(x^{\mu }) \) are the
standard left-handed quark doublets and right-handed singlets, respectively,
which live in four dimensions. They carry additional flavor and color indices
which have been suppressed. \( Y_{u} \) are \( 3\times 3 \) matrices in the
flavor space. We have not written the Yukawa interaction of the down quarks
because it is proportional to the down quark masses which are small. Of course
these interactions are present and necessary for generating down quark masses
and mixings. We have also omitted the kinetic terms of fermions, as well as
gauge bosons interactions, which will not be relevant in our approximation.
The rôle of \( \delta (x^{5}) \) is to force the fermions to live in four dimensions.
As usual, one assumes that the fifth dimension \( x^{5} \) is compactified
on a circle of radius \( R \) with the points \( x^{5} \) and \( -x^{5} \)
identified (that is, an orbifold \( S^{1}/{\mathbb Z}_{2} \)). Fields even
under the \( {\mathbb Z}_{2} \) symmetry will have zero modes which will be
present in the low energy theory. Fields odd under \( {\mathbb Z}_{2} \) will
only have KK modes and will disappear from the low energy spectrum. One chooses
the scalar doublet to be even under the \( {\mathbb Z}_{2} \) symmetry in order
to have a standard zero mode Higgs field. Following the standard Kaluza-Klein
construction, we Fourier expand the scalar fields as follows (from now on \( x \)
refers only to the four dimensional coordinates \( x^{\mu } \)) \begin{equation}
\varphi (x^{\mu },x^{5})=\sum _{n=0}^{\infty }\cos \frac{nx^{5}}{R}\varphi _{n}(x^{\mu })\, .
\end{equation}
 Substituting this expression into the fifth dimensional Lagrangian, eq.~\prettyref{eq:5d-lagrangian},
and integrating over the fifth component leads to the four dimensional Lagrangian
for the KK modes \( \varphi _{n}(x) \). The kinetic terms, however, are not
canonical, and we need to perform the following redefinitions of fields and
couplings in order to cast them into canonical form:

\begin{equation}
\varphi _{0}(x)\rightarrow \frac{1}{\sqrt{2\pi R}}\varphi _{0}(x)\: ,\quad \varphi _{n}(x)\rightarrow \frac{1}{\sqrt{\pi R}}\varphi _{n}(x)\: ,\; (n\not =0)\: ,\quad Y_{u}\rightarrow \sqrt{2\pi R}\, Y_{u}\, .
\end{equation}
 Then, we arrive at the following four dimensional Lagrange density \begin{eqnarray}
\mathcal{L} & = & \partial _{\mu }\varphi _{0}^{\dagger }\partial ^{\mu }\varphi _{0}-\left( \bar{Q}_{L}Y_{u}u_{R}\varphi _{0}+\textrm{h}.\textrm{c}.\right) \nonumber \\
 & + & \sum _{n=1}^{\infty }\left( \partial _{\mu }\varphi _{n}^{\dagger }\partial ^{\mu }\varphi _{n}-\frac{n^{2}}{R^{2}}\varphi _{n}^{\dagger }\varphi _{n}-\left( \bar{Q}_{L}Y_{u}u_{R}\sqrt{2}\varphi _{n}+\textrm{h}.\textrm{c}.\right) \right) \, ,\label{eq:lagrangian4d} 
\end{eqnarray}
 which will be used in our calculations. Fermions obtain their masses when the
neutral component of the zero mode Higgs field, \( \varphi ^{(0)}_{0} \), acquires
a vacuum expectation value \( \left\langle \varphi ^{(0)}_{0}\right\rangle \equiv v \).
Mass matrices are diagonalized in the standard way, and, if we only keep terms
proportional to the top quark mass we obrtain the following Yukawa interaction
between the mass eigenstates and the KK modes of scalar fields \begin{equation}
\label{eq:yukawa}
{\mathcal{L}}_{Y}=-\sqrt{2}\frac{m_{t}}{v}\sum ^{\infty }_{n=1}\left( \overline{t}_{L}\, t_{R}\varphi ^{(0)}_{n}+\sum _{f}^{d,s,b}\overline{f}_{L}V^{*}_{tf}\, t_{R}\varphi ^{(-)}_{n}+\mathrm{h}.c.\right) \, ,
\end{equation}
 where \( V_{tf} \) is the Cabibbo-Kobayashi-Maskawa (CKM) matrix, while \( \varphi ^{(0)}_{n} \)
and \( \varphi ^{(-)}_{n} \) are the neutral and charged components of the
KK scalar doublets, respectively. Notice the additional factor \( \sqrt{2} \)
in the coupling of the KK modes, which comes from the normalization of the zero
mode in the Fourier expansion. 

\begin{figure}
{\par\centering \includegraphics{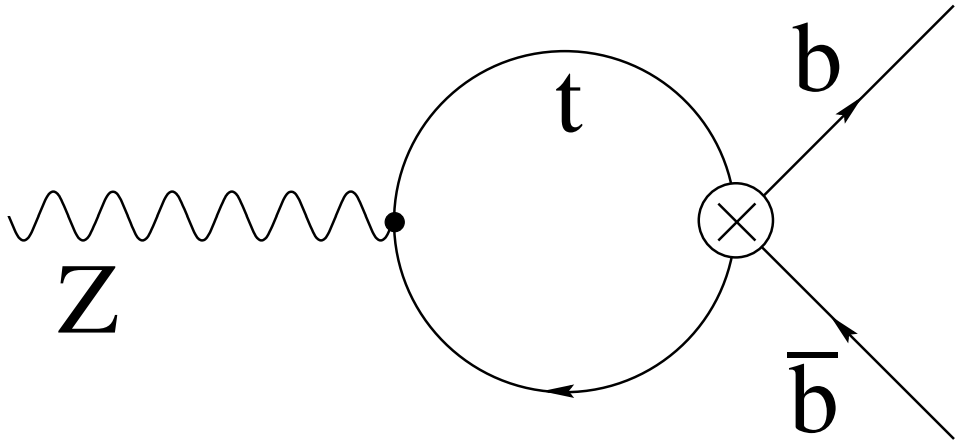} \par}

\caption{Effective field theory diagram used in the computation of the leading logarithmic
corrections induced by four fermion interactions.\label{fig:extrabb-eff}}
\end{figure}

In the low energy limit one can integrate out the KK modes (by using the equations
of motion, for instance) and obtain the following four fermion interaction (in
the weak basis)

\begin{equation}
\label{eq:lagrangian4f}
{\mathcal{L}}_{\mathrm{eff}}=\frac{(\pi R)^{2}}{3}(\bar{Q}_{L}Y_{u}u_{R})(\bar{u}_{R}Y_{u}^{\dagger }Q_{L})\, ,
\end{equation}
 which can be expressed in terms of the mass eigenstates (keeping only terms
proportional to \( m_{t}). \)

\begin{equation}
\label{eq:lagrangian-diagonal}
{\mathcal{L}}_{\mathrm{eff}}=\frac{(\pi R)^{2}}{3}\frac{m^{2}_{t}}{v^{2}}\left[ \left( \overline{t}_{L}\, t_{R}\right) \left( \overline{t}_{R}\, t_{L}\right) +\sum ^{d,s,b}_{f,f'}\left( \overline{f'}_{L}\, t_{R}\right) \left( \overline{t}_{R}\, f_{L}\right) \, V_{tf}V^{*}_{tf'}\right] \, .
\end{equation}
 The above Lagrangian provides, for instance, a four fermion interaction \( (\bar{b}_{L}\, t_{R})(\bar{t}_{R}\, b_{L}) \)
(and also \( (\bar{s}_{L}\, t_{R})(\bar{t}_{R}\, s_{L}) \) and \( (\bar{d}_{L}\, t_{R})(\bar{t}_{R}\, d_{L}) \)
) which much in the spirit of ref.~\cite{bernabeu:1997zh} can contribute at
one loop level to the decay \( Z\rightarrow b\bar{b} \) as well to \( B \)-\( \bar{B} \)
and \( K \)-\( \bar{K} \) mixing. However, if we use the effective four-fermion
interaction, the loop integral in fig.~\ref{fig:extrabb-eff} is divergent
and following ref.~\cite{bernabeu:1997zh} one can only compute in this way
the dominant logarithmic contributions. To obtain the non-logarithmic parts
one should calculate the one-loop matching with the complete theory. One of
the advantages of models with large extra dimensions is that they provide this
full theory, which will allow us, as we will immediately show, to compute not
only the logarithmic corrections but also the finite parts. In order to accomplish
this, one has to maintain all KK modes as dynamical particles. Therefore, in
what follows we will use the interactions given in eq.~\prettyref{eq:yukawa}.

\section{\protect\( Z\rightarrow \lowercase {b\bar{b}}\protect \) }

\label{sec:zbb}
\begin{figure}
{\par\centering \includegraphics{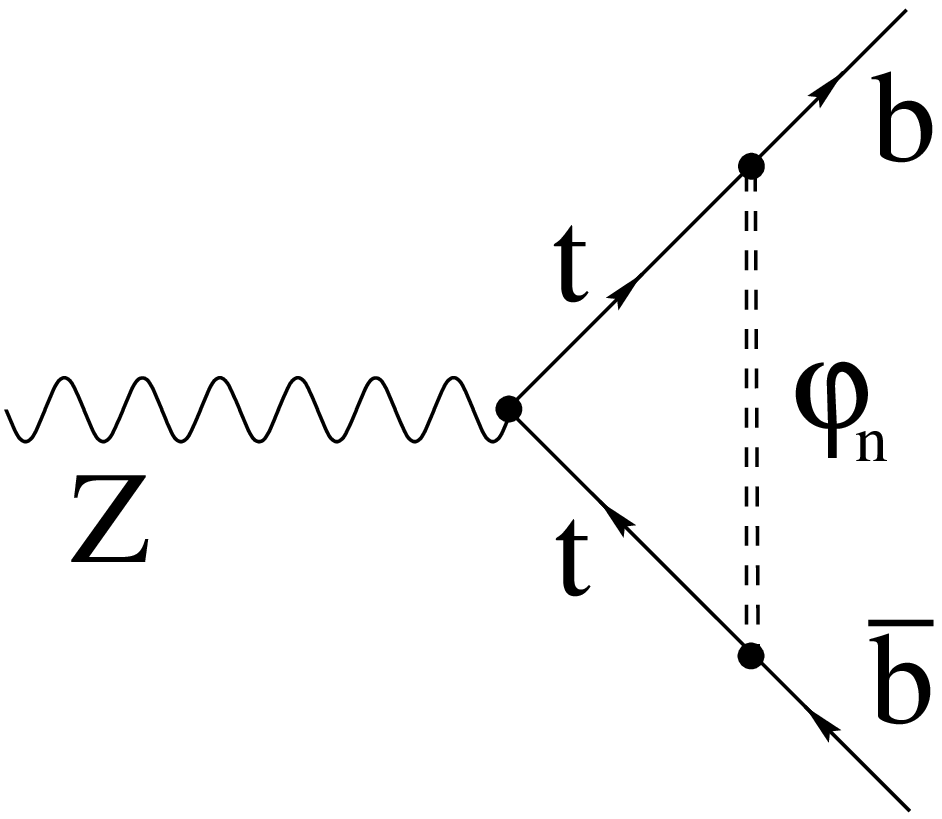} \par}

\caption{Diagram contributing to \protect\( Z\rightarrow b\bar{b}\protect \) if the
scalar doublet lives in five dimensions. The tower of KK modes of charged scalars
is represented by the dashed double line.\label{fig:extrabb}}
\end{figure}

In the SM there are many diagrams contributing to the vertex
corrections to \( Z\rightarrow b\bar{b} \). In the Feynman or in the Landau
gauges the dominant contribution for large \( m_{t} \) is captured by diagrams
such as the one shown in fig.~\ref{fig:extrabb}, with a charged would-be Goldstone
boson running in the loop. In the unitary gauge these corrections originate
from the longitudinal parts of the gauge boson propagators. In general there
are strong cancellations among vertex diagrams (as the graph of fig.~\ref{fig:extrabb})
and diagrams with self-energies in the external fermion legs in such a way that
the dominant contribution is finite. By far the easiest way to compute these
corrections is to resort to the equivalence theorem \cite{cornwall:1974km,vayonakis:1976vz,chanowitz:1985hj,gounaris:1986cr},
i.e. to use the Ward identities \cite{lytel:1980zh,barbieri:1992nz,barbieri:1993dq,papavassiliou:1996hj}
that relate the \( Z \)-\( b\bar{b} \) vertex to the \( G^{0} \)-\( b\bar{b} \)
vertex , where \( G^{0} \) denotes the would-be Goldstone boson associated
to the \( Z \) gauge boson. 

In the model we are considering, there are additional contributions enhanced
by \( m_{t} \) that arise from the presence of the charged scalar KK modes,
with interactions governed by eq.~\prettyref{eq:yukawa}, which give rise to
the diagram of fig.~\ref{fig:extrabb}. If the gauge bosons also possess KK
modes there will be additional diagrams, such as the one shown in fig.~\ref{fig:extrabb},
in which the KK modes of the scalars will be replaced by the corresponding KK
modes of the \( W \)-gauge bosons. Even though their contribution is formally
suppressed by a factor \( (m_{W}/m_{t})^{2} \), we will estimate it at the
end of this section. In such a case it is important to realize that the KK modes
of the charged scalars appearing inside the loop are not the would-be Goldstone
bosons of the KK modes of the gauge bosons. In fact the mass of the KK modes
associated to the gauge bosons is given by their fifth components\footnote{%
Fifth components of gauge bosons are odd under the \( {\mathbb Z}_{2} \) symmetry,
therefore they do not have zero modes. Masses for the zero-mode gauge bosons
should be provided by the usual Higgs mechanism, while masses for non-zero mode
gauge bosons are provided by their fifth components. Thus, only the zero mode
scalars play the rôle of Goldstone bosons.
}. This distinction becomes clear if one uses the unitary gauge for the KK modes
of the gauge bosons. In this case the fifth components of the five dimensional
gauge bosons are completely absorbed by the KK modes of the gauge bosons, i.e.
there are not graphs containing would-be Goldstone bosons, while the KK modes
of the scalars remain in the spectrum of physical particles, i.e. the diagram
of fig.~\ref{fig:extrabb} persists.

Again, the easiest way to calculate the contribution of the scalar KK modes
is to resort to the equivalence theorem for the external \( Z \) and compute
the diagram of fig.~\ref{fig:extrabb} with the Z replaced by the \( G^{(0)} \).
Since the couplings of the KK modes to fermions are universal, summing all scalar
contributions amounts to replacing the propagator of the SM would-be Goldstone
boson by (for Euclidean momenta, which we will use in the momentum integrals
after the Wick rotation)

\begin{equation}
\label{eq:prop-minkowski}
\frac{1}{k_{E}^{2}}\rightarrow \frac{1}{k_{E}^{2}}+2\sum _{n=1}^{\infty }\frac{1}{k_{E}^{2}+n^{2}/R^{2}}=\sum _{n=-\infty }^{\infty }\frac{1}{k_{E}^{2}+n^{2}/R^{2}}=\pi R\frac{\coth (k_{E}\pi R)}{k_{E}},
\end{equation}
where \( k_{E}=\sqrt{k^{2}_{E}}. \)

Notice the behavior of this propagator: for small \( k_{E} \) it reduces to
the standard Goldstone propagator plus, if expanded at leading order, an additional
constant which furnishes the contact interaction derived above, eq.~\prettyref{eq:lagrangian4f}.
However, for large \( k_{E} \) it goes as \( 1/k_{E} \); as a result the ultraviolet
(UV) behavior of this theory is worse than in the SM by one power of \( k_{E} \),
a fact which will eventually trigger the non-renormalizability of the theory.
However, since in the large \( k_{E} \) limit only even powers of \( k_{E} \)
contribute in standard QFT integrals, this worse UV behavior of the non-standard
propagator does not create additional problems, as long as only one such propagator
is inserted into a convergent graph. For instance, the dominant SM contribution
to the \( Zb\bar{b} \) vertex, fig.~\ref{fig:extrabb}, is convergent because
the integrand behaves as \( 1/k_{E}^{6}; \) when we use the non-standard propagator
this behavior will drop to \( 1/k_{E}^{5} \), which still leads to a convergent
result.

To see how this works in detail we parametrize the effective \( Z\bar{b}b \)
vertex as 

\begin{equation}
\frac{g}{c_{W}}\, \bar{b}\gamma ^{\mu }(g_{L}P_{L}+g_{R}P_{R})bZ_{\mu }\, ,
\end{equation}
where \( P_{L}=(1-\gamma _{5})/2 \) and \( P_{R}=(1+\gamma _{5})/2 \) are,
respectively, the left and right chirality projectors, \( c^{2}_{W}=1-s^{2}_{W}=m_{W}^{2}/m^{2}_{Z} \),
and \begin{eqnarray}
g_{L} & = & -\frac{1}{2}+\frac{1}{3}s_{W}^{2}+\delta g^{\mathrm{SM}}_{L}+\delta g^{\mathrm{NP}}_{L}\nonumber \\
g_{R} & = & \frac{1}{3}s_{W}^{2}+\delta g^{\mathrm{SM}}_{R}+\delta g^{\mathrm{NP}}_{R}\, .
\end{eqnarray}
In the above equations the \( -1/2+s_{W}^{2}/3 \) and \( s_{W}^{2}/3 \) are
the tree level contributions, \( \delta g^{\mathrm{SM}}_{L} \) and \( \delta g^{\mathrm{SM}}_{L} \)
denote higher order corrections within the SM, whereas \( \delta g^{\mathrm{NP}}_{L} \)
and \( \delta g^{\mathrm{NP}}_{L} \) parametrize the contributions coming from
new physics. Notice that, in general, \( g_{R} \) only receives sub-dominant
corrections (not proportional to the top quark mass) in both the SM and in most
of new physics scenarios. In particular, the dominant SM contribution comes
from the Goldstone boson diagrams running in the loop, fig.~\ref{fig:extrabb},
and it is given by \begin{equation}
\label{eq:deltasm}
\delta g_{L}^{\mathrm{SM}}\approx \sqrt{2}G_{F}m_{t}^{4}\, i\int \frac{d^{4}k}{(2\pi )^{4}}\frac{1}{(k^{2}-m_{t}^{2})^{2}k^{2}}=\frac{\sqrt{2}G_{F}m_{t}^{2}}{(4\pi )^{2}}\, .
\end{equation}
 Therefore, adding the KK modes we obtain \begin{equation}
\label{eq:deltanp}
\delta g^{\mathrm{NP}}_{L}\approx \delta g_{L}^{\mathrm{SM}}\left( F(a)-1\right) \, ,
\end{equation}
 where \( a=\pi Rm_{t} \), and \begin{eqnarray}
F(a)\,  & = & \left. \pi R\int \frac{dk_{E}k^{2}_{E}\coth (k_{E}\pi R)}{(k^{2}_{E}+m^{2}_{t})^{2}}\right/ \int \frac{dk_{E}k_{E}}{(k^{2}_{E}+m^{2}_{t})^{2}}\nonumber \\
 & = & 2a\int _{0}^{\infty }dx\frac{x^{2}}{(1+x^{2})^{2}}\coth (ax)\label{eq:integral} 
\end{eqnarray}
 is the ratio of the non-standard to the standard integrals (in the Euclidean).
\( F(a) \) is, as expected, perfectly convergent. It can be expanded for small
\( a \), yielding \begin{equation}
\label{eq:f-function}
F(a)\approx 1+a^{2}\left( -\frac{1}{3}-\frac{4}{\pi ^{2}}\zeta '(2)-\frac{2}{3}\log (a/\pi )\right) \approx 1+a^{2}\left( 0.80979-\frac{2}{3}\log (a)\right) \, ,
\end{equation}
 where \( \zeta ' \) is the derivative of the Riemann zeta function. As commented
before, the logarithmic contribution can be obtained easily by using the four-fermion
interaction at the loop level and then cutting off the integrals at \( k_{E}\approx 1/a \).
This model, in addition to the logarithmic contribution provides also the non-logarithmic
piece, and the result reported above is valid for any value of \( a \). One
important point about this result is that the additional contribution from the
KK modes is always positive, a fact which will be of particular importance in
the following phenomenological analysis.

A shift in the \( Zb\bar{b} \) couplings gives a shift in \( R_{b}=\Gamma _{b}/\Gamma _{h} \)
(here \( \Gamma _{b}=\Gamma (Z\rightarrow b\bar{b}) \) and \( \Gamma _{h}=\Gamma (Z\rightarrow \mathrm{hadrons}) \))
given by \begin{equation}
\label{eq:rb}
R_{b}=R_{b}^{\mathrm{SM}}\frac{1+\delta ^{\mathrm{NP}}_{bV}}{1+R_{b}^{\mathrm{SM}}\delta ^{\mathrm{NP}}_{bV}},
\end{equation}
 where \begin{equation}
\label{eq:deltavnp}
\delta ^{\mathrm{NP}}_{bV}=\frac{\delta \Gamma _{b}}{\Gamma _{b}^{\mathrm{SM}}}\approx 2\frac{g_{L}}{(g_{L})^{2}+(g_{R})^{2}}\delta g_{L}^{\mathrm{NP}}\approx -4.6\, \delta g_{L}^{\mathrm{NP}}
\end{equation}
 gives the relative change to \( \Gamma _{b} \) due to vertex corrections coming
from new physics, \( \Gamma _{b}=\Gamma ^{\mathrm{SM}}_{b}+\delta \Gamma _{b} \).
Here, quantities with superscript SM denote standard model values including
complete radiative corrections. Note that non-vertex corrections are universal
for all quarks and cancel in the ratio \( R_{b} \).

In recent years there has been a significant controversy surrounding \( R_{b} \),
because for some time its measured value was more than two standard deviations
away from the one predicted in the SM. However, the present experimental value
is perfectly compatible with the SM \cite{groom:1998in}: \( R_{b}^{\mathrm{exp}}=0.2164\pm 0.00073 \),
while \( R_{b}^{\mathrm{SM}}=0.2157\pm 0.0002 \), although the central value
is still somewhat higher. Using these values together with equations~\prettyref{eq:deltasm},
\prettyref{eq:deltanp}, \prettyref{eq:rb} and \prettyref{eq:deltavnp}, one
immediately finds that \( F(a)-1=-0.24\pm 0.31 \). However, as commented before,
\( F(a) \) is always larger than \( 1 \) since corrections from extra dimensions
are always positive. In this case one should be especially careful when estimating
confidence levels (CL) for the bounds on \( F(a)-1 \). For this purpose we
used the prescription of ref.~\cite{feldman:1998qc}, which provides more reliable
limits than other approaches, and found the following 95\% CL limit of \( F(a)-1<0.39 \).
After evaluation of the integral \prettyref{eq:integral} the previous limit
translates into an upper bound on \( a \), \( a<0.56 \), which amounts to
the following lower bound on the compactification scale \( M_{c} \),\begin{equation}
M_{c}>1\: \mathrm{TeV}\, .
\end{equation}
If only 68\% CL limits are required we obtain \( F(a)-1<0.11 \), \( a<0.26 \),
and \( M_{c}>2\: \mathrm{TeV} \). Quite interestingly, these one-loop bounds
are comparable to those obtained from tree level processes \cite{masip:1999mk,delgado:1999sv,rizzo:1999br,carone:1999nz,nath:1999fs,nath:1999mw}. 

In the above discussion we have not taken into account the effects of the gauge
boson KK modes because their contribution is suppressed by \( (m_{W}/m_{t})^{2} \).
However, since \( (m_{W}/m_{t})^{2}\sim 1/4 \), such contributions, even though
formally suppressed, could become numerically relevant and affect the obtained
bounds. In addition, those contributions are present even if the scalar doublet
lives in four dimensions and, as a consequence, has no KK modes. Therefore,
we will provide an estimate of their size. 

At energies below the compactification scale one can integrate the KK modes
of the gauge bosons and obtain the following four-fermion interaction for the
third generation (in the eigenvalue basis and neglecting CKM mixings)

\begin{equation}
\label{eq:gauge-4fermion}
{\mathcal{L}}_{\mathrm{gauge}}=-\frac{(\pi R)^{2}}{3}\frac{g^{2}}{2}\left( \bar{b}_{L}\gamma _{\mu }t_{L}\right) \left( \bar{t}_{L}\gamma ^{\mu }b_{L}\right) \, ,
\end{equation}
to be compared with the contribution from scalar modes obtained from \prettyref{eq:lagrangian-diagonal}
(again neglecting CKM mixings)\begin{equation}
\label{eq:yukawa-4fermion}
{\mathcal{L}}_{\mathrm{Ytb}}=\frac{(\pi R)^{2}}{3}\frac{m^{2}_{t}}{v^{2}}\left( \bar{b}_{L}\, t_{R}\right) \left( \bar{t}_{R}\, b_{L}\right) \, .
\end{equation}
 As commented above and discussed in ref.~\cite{bernabeu:1997zh}, one can
use these effective Lagrangians to obtain the leading logarithmic corrections
to \( Z\rightarrow b\bar{b} \). In order to achieve that we compute the divergent
part of the diagram shown in fig.~\ref{fig:extrabb-eff}, where the symbol
\( \otimes  \) denotes the insertion of any of these four-fermion operators.
It turns out that the different Lorentz structure of the two four fermion interactions
in \prettyref{eq:gauge-4fermion} and \prettyref{eq:yukawa-4fermion} gives
an additional factor \( -2 \) in the former case. Therefore, up to non-logarithm
corrections, one can include the effect of the exchange of KK modes of gauge
bosons by multiplying the effect of the scalar KK modes by a factor \( 1+2(m_{W}/m_{t})^{2} \),
which gives a non-negligible correction. Notice that due to the positive relative
sign, inclusion of this correction would lead to a 20\% improvement in the bound
on \( M_{c}. \) Moreover, this correction will remain even in the absence of
scalar KK modes; in that case one can still place a bound on \( M_{c} \) of
about \( 0.7 \)~TeV.

\section{Box contributions to \protect\( K\protect \)-\protect\( \bar{K}\protect \)
and \protect\( B\protect \)-\protect\( \bar{B}\protect \) mixing and the divergences}

\label{sec:box}In the SM, the mixing between the \( B^{0} \) meson and its
anti-particle is also completely dominated by the top-quark contribution. The
explicit \( m_{t} \) dependence of the box diagrams is given by the loop function\cite{buchalla:1996vs}\begin{equation}
S(x_{t})_{\textrm{SM}}=\frac{x_{t}}{4}\left[ 1+\frac{9}{1-x_{t}}-\frac{6}{(1-x_{t})^{2}}-\frac{6x_{t}^{2}\log (x_{t})}{(1-x_{t})^{3}}\right] \; ,\qquad x_{t}\equiv \frac{m^{2}_{t}}{M_{W}^{2}},
\end{equation}
 which contains the hard \( m_{t}^{2} \) term, i.e. \( x_{t}/4 \), induced
by the longitudinal \( W \) exchanges. The same function controls the top--quark
contribution to the \( K \)-\( \bar{K} \) mixing parameter \( \varepsilon _{K} \).
The measured top-quark mass, \( m_{t}=175 \)~GeV, implies \( S(x_{t})_{\textrm{SM}}\sim 2.5 \).

\begin{figure}
{\par\centering \includegraphics{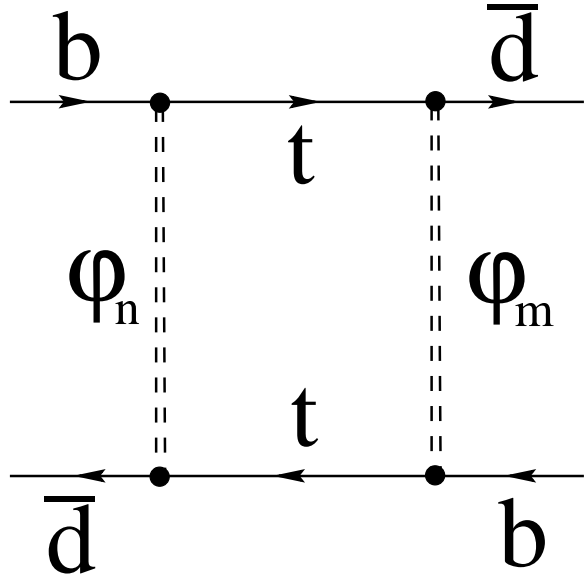} \par}

\caption{Box diagram contributing to \protect\( B\protect \)-\protect\( \bar{B}\protect \)
and \protect\( K\protect \)-\protect\( \bar{K}\protect \) mixings. The tower
of KK modes is represented by the dashed double lines.\label{fig:extrabox}}
\end{figure}

The KK modes of the charged components of the doublet also contribute to this
box diagram. The total dominant contribution, SM plus KK modes, can be obtained
by substituting the propagator~\prettyref{eq:prop-minkowski} in the box diagram,
fig.~\ref{fig:extrabox}. However, as discussed in the previous section, the
modified propagator behaves as \( 1/k_{E} \) for large \( k_{E} \), and therefore,
the insertion of two propagators of this type turns this modified diagram into
UV divergent. On the other hand, the insertion of only one modified propagator
still yields a finite result. 

We write the correction to \( S(x_{t}) \) as 

\begin{equation}
S(x_{t})=S(x_{t})_{\textrm{SM}}+\delta S(x_{t})\, ,\qquad \quad \delta S(x_{t})=\frac{x_{t}}{4}\, \left( G(a)-1\right) \, \, \, \, ,
\end{equation}
 where the function \( G(a) \) is again the ratio of the non-standard to standard
box integrals\footnote{%
Notice that, even though the SM box integral is given exactly by the same expression
as that of the SM vertex integral in the previous section, their original structures
are rather different . In particular, the box diagram contains two scalar propagators
whereas the vertex diagram only contains one.
}

\begin{eqnarray}
G(a) & = & \left. (\pi R)^{2}\int \frac{dk_{E}k^{3}_{E}\coth ^{2}(k_{E}\pi R)}{(k^{2}_{E}+m^{2}_{t})^{2}}\right/ \int \frac{dk_{E}k_{E}}{(k^{2}_{E}+m^{2}_{t})^{2}}\nonumber \\
 & = & 2a^{2}\int _{0}^{\infty }dx\frac{x^{3}}{(1+x^{2})^{2}}\coth ^{2}(ax)\label{eq:integral-box} 
\end{eqnarray}
 which is clearly divergent for \( x\rightarrow \infty  \). In order to estimate
this integral, we split \( \coth (ax)\rightarrow \frac{1}{ax}(1+\underbrace{ax\coth (ax)-1}) \)
and rewrite \( G(a) \) as \begin{eqnarray}
G(a) & = & 2\int _{0}^{\infty }dx\frac{x}{(1+x^{2})^{2}}\left( 1+2\left( ax\coth (ax)-1\right) +\left( ax\coth (ax)-1\right) ^{2}\right) \nonumber \\
 & = & 1+2\left( F(a)-1\right) +2\int _{0}^{\infty }dx\frac{x}{(1+x^{2})^{2}}\left( ax\coth (ax)-1\right) ^{2}\, .\label{eq:g-funct} 
\end{eqnarray}
The divergence is contained in the last term. To evaluate it we cut off the
integral at \( x\approx n_{s}/a \), where \( n_{s} \) is related to the scale
at which new physics enters to regulate the five dimensional theory. In particular,
\( M_{s}\sim n_{s}M_{c} \) and \( n_{s}\gg 1 \). Then, after a change of variable
\( y=ax \) the last term can be re-written as\begin{eqnarray}
2a^{2}\int _{0}^{n_{s}}dy\frac{y}{(a^{2}+y^{2})^{2}}\left( y\coth (y)-1\right) ^{2} & \approx  & 2a^{2}\int _{0}^{n_{s}}dy\frac{1}{y^{3}}\left( y\coth (y)-1\right) ^{2}\nonumber \\
 & \approx  & 2a^{2}\left( -1.38136+\log (n_{s})\right) \, ,\label{eq:div-box} 
\end{eqnarray}
where in the second expression we have assumed \( a\ll 1 \), and, in addition,
in the last expression we have also taken \( n_{s}\gg 1 \). Combining this
result with \prettyref{eq:g-funct} and \prettyref{eq:f-function} we obtain
\begin{equation}
G(a)\approx 1+a^{2}\left( -1.14314-\frac{4}{3}\log (a)+2\log (n_{s})\right) .
\end{equation}
We have checked that the coefficients of the two logarithms, \( \log (a) \)
and \( \log (n_{s}) \), can also be obtained by performing first the convergent
momentum integrals and subsequently truncating the divergent double series at
\( \sim n_{s} \). However, this latter method is technically far more complicated
than the one presented here. 

For moderate values of \( a\sim 0.2 \) and \( n_{s}\sim 10 \) the new physics
correction is only about \( 0.2 \). For more extreme values (for instance \( a\sim 0.6 \)
and \( n_{s}\sim 100 \)), we find that the contribution from extra dimensions
to the function \( G(a) \) is about \( 3 \). Notice also that, \underbar{}as
discussed at the end of sec.~\ref{sec:zbb}, the presence of diagrams with
gauge boson KK modes could modify the bounds on \( M_{c} \) by a factor of
about 20\%. However, given the uncertainty in the calculation of the box diagrams
due to the dependence on the scale \( M_{s}, \) estimating such effects seems
superfluous. The important point, however, is that the contribution from extra
dimensions to the function \( S(x_{t}) \) is always positive.

We can use the measured \( B^{0}_{d} \)-\( \bar{B}^{0}_{d} \) mixing to infer
the experimental value of \( S(x_{t}) \) and, therefore, to set a limit on
the \( \delta S(x_{t}) \) contribution. The explicit dependence on the quark--mixing
parameters can be resolved by combining the constraints from \( \Delta M_{B^{0}_{d}} \),
\( \varepsilon _{K}, \) and \( \Gamma (b\rightarrow u)/\Gamma (b\rightarrow c) \).
In ref.~\cite{bernabeu:1997zh} a complete analysis of the allowed values for
\( S(x_{t}) \) was performed by varying all parameters in their allowed regions.
The final outcome of such an analysis is that \( S(x_{t}) \) could take values
within a rather large interval, namely 

\begin{equation}
1<S(x_{t})<10\, .
\end{equation}
Since most of the errors come from uncertainties in theoretical calculations,
it is rather difficult to assign confidence levels to the bounds quoted above.
The lower limit is very stable under changes of parameters, while the upper
limit could be modified by a factor of 2 by simply doubling some of the errors. 

Given that the standard model value for \( S(x_{t}) \) is \( S(x_{t})_{\mathrm{SM}}=2.5 \),
\emph{positive} contributions can be comfortably accommodated, whereas negative
contributions are more constrained. As we have seen, extra dimensions result
in \emph{positive} contributions to \( S(x_{t}) \); in fact one can obtain
values that could approach the upper limit of \( S(x_{t}) \) only for rather
small values of the compactification scale \( M_{c} \) and large values of
the scale of new physics, \( M_{s} \). It seems therefore that, at present,
the above bounds do not provide good limits on \( M_{c} \). On the other hand,
if future experiments combined with theoretical improvements were to furnish
a value for \( S(x_{t}) \) exceeding that of the SM, our analysis shows that
such a discrepancy could easily be accommodated in models with large extra dimensions.

\section{Conclusions}

\label{sec:conclusions}We have studied, at the one loop level, the minimal
extension of the SM with one extra dimension compactified in \( S^{1}/{\mathbb Z}_{2} \).
Fermions live in 4 dimensions, while gauge bosons and the scalar doublet live
in 5 dimensions and therefore give rise to a tower of KK modes. In the case
of a single extra dimension the contribution of the infinite tower of KK modes
lead to finite tree level predictions. We have investigated whether this feature
persists at the one loop level, by considering two amplitudes which are enhanced
by the top-quark mass, namely \( Z\rightarrow b\bar{b} \) and \( B \)-\( \bar{B} \)
mixing. 

The infinite tower of KK modes enters in the calculation of \( Z\rightarrow b\bar{b} \)
by modifying the propagator of the charged scalars running in the vertex diagram.
This can be effectively taken into account by using a modified propagator for
the scalars which for large \( k \) behaves as \( 1/k \), instead of the canonical
behavior of \( 1/k^{2} \). In spite of that the effect is finite and calculable.
The result, when compared with precise experimental data on \( R_{b}=\Gamma _{b}/\Gamma _{h} \),
is used to place stringent limits on the compactification scale, \( M_{c}, \)
\( M_{c}>1 \)~TeV at the 95\% CL, which are comparable to the bounds obtained
from tree-level processes. 

The box diagrams contributing to \( B \)-\( \bar{B} \) and \( K \)-\( \bar{K} \)
mixings contain two propagators of KK modes. The double sum over KK modes amounts
to the replacement of both propagators by the aforementioned softer ones, a
fact which increases the UV behavior of the diagram by two powers, and renders
it divergent. Thus, due to such contributions the theory becomes non-renormalizable
already at the one-loop level. To estimate their size one has to assume that
the model is embedded in a more complete theory which would provide an effective
cutoff at scales larger than \( M_{c} \). In practice, this can be realized
either by cutting off the infinite integrals at momenta of order of \( M_{s} \),
the scale where new physics enters to regularize the five dimensional theory,
or by truncating the sum of KK modes at some value of \( n_{s} \), \( n_{s}\sim M_{s}/M_{c} \),
with \( n_{s} \) expected to be order \( \sim 100 \) or less. This way we
can estimate the correction induced by the extra dimension to the function \( S(x_{t}) \)
which parametrizes the short distance physics in \( B \)-\( \bar{B} \) and
\( K \)-\( \bar{K} \) mixings. A phenomenological analysis shows that \( 1<S(x_{t})<10, \)
while the SM value is \( S(x_{t})=2.5 \). This suggests that moderate \emph{positive}
extra contributions to \( S(x_{t}) \) are still allowed. Since within the model
we consider the contributions to \( S(x_{t}) \) from KK modes is always positive
and moderate in size, no interesting bounds can be obtained form this process.
However, if in the future a value of \( S(x_{t}) \) larger than the SM value
is found, extra dimensions could easily accommodate it.

\acknowledgements

This work has been funded by CICYT under the Grant AEN-99-0692, by DGESIC under
the Grant PB97-1261 and by the DGEUI of the Generalitat Valenciana under the
Grant GV98-01-80.

\end{document}